\documentclass[english]{emulateapj}
\usepackage{lmodern}

\usepackage[T1]{fontenc}
\usepackage[latin9]{inputenc}
\usepackage{color}
\usepackage{array}
\usepackage{float}
\usepackage{dvipost}
\usepackage{multirow}
\usepackage{amssymb}
\usepackage{graphicx}
\usepackage{esint}

\makeatletter

\providecommand{\tabularnewline}{\\}
\dvipostlayout
\dvipost{osstart color push Red}
\dvipost{osend color pop}
\dvipost{cbstart color push Blue}
\dvipost{cbend color pop}

\makeatother

\usepackage{babel}
\begin{document}

\title{The impact of mass segregation and star-formation on the rates \\
of gravitational-wave sources from extreme mass ratio inspirals }

\author{Danor Aharon \& Hagai B. Perets}

\affil{Physics Department, Technion - Israel Institute of Technology, Haifa,
Israel 3200003}
\begin{abstract}
Compact stellar objects inspiralling into massive black holes (MBHs)
in galactic nuclei are some of the most promising gravitational wave
(GWs) sources for next generation GW-detectors. The rates of such
extreme mass ratio inspirals (EMRIs) depend on the dynamics and distribution
of compact objects around the MBH. Here we study the impact of mass-segregation
processes on EMRI rates. In particular, we provide the expected mass
function of EMRIs, given an initial mass function of stellar BHs (SBHs),
and relate it to the mass-dependent detection rate of EMRIs. We then
consider the role of star formation on the distribution of compact
objects and its implication on EMRI rates. We find that the existence
of a wide spectrum of SBH masses lead to the overall increase of EMRI
rates, and to high rates of the EMRIs from the most-massive SBHs.
However, it also leads to a relative quenching of EMRI rates from
lower-mass SBHs, and together produces a steep dependence of the EMRI
mass function on the highest-mass SBHs. Star-formation history plays
a relatively small role in determining the EMRI rates of SBHs, since
most of them migrate close to the MBH through mass-segregation rather
than forming in-situ. However, the EMRI rate of neutron stars can
be significantly increased when they form in-situ close to the MBH,
as they can inspiral before relaxation processes significantly segregates
them outwards. A reverse but weaker effect of decreasing the EMRI
rates from neutron stars and white dwarfs occurs when star-formation
proceeds far from the MBH. 
\end{abstract}

\section{INTRODUCTION}

Nuclear stellar clusters, (NSCs) hosting massive black holes (MBHs)
are thought to exist in a significant fraction of all galactic nuclei,
including our own \citep{Eisenhauer2005,2005ApJ...620..744G}. The
dynamics of stars in the dense NSC lead to strong interactions between
stars and the MBH. In particular compact objects (COs) such as white
dwarfs (WDs), neutron stars (NSs), and stellar black holes (SBHs),
may inspiral onto the MBH and emit gravitational waves (GWs) observable
to cosmological distances with next generation GW detectors. The inspiral
of a CO into an MBH (\textquotedblleft extreme mass ratio inspiral\textquotedblright{}
{[}EMRI{]}) is among the main targets of future Evolved Laser Interferometer
Space Antenna (eLISA). 

The properties of EMRIs and their rates depend strongly on the evolution
and dynamics of different stellar populations near MBHs, processes
in which mass segregation processes play a key role (see \citealt[keshet][]{2006ApJ...649...91F,2006ApJ...645L.133H,2009ApJ...697.1861A,2009ApJ...698L..64K,Preto2010}
and \citet{Ama+11} for a detailed treatment of mass segregation near
MBHs). Mass segregation occurs through two-body interactions between
less massive and more massive objects. The encounters drive stellar
populations of different masses towards energy equipartition which
results in the more massive objects migrating closer to the center,
while the less massive ones migrate outwards. In particular, mass
segregation can increase the density of the more massive COs within
the region $r<a_{GW}$, where $a_{GW}$ is the maximal semi-major
axis at which a CO could still inspiral and become an eLISA source
(hereafter the critical separation), rather than plunge-in on a too
radial orbit and spending too short time in the GW detector band and
unlikely to be detected. \citet{2005ApJ...629..362H} derived an analytical
order-of-magnitude estimate for the critical separation, given by
\begin{equation}
a_{GW}=r_{h}\left(\frac{d_{c}}{r_{h}}\right)^{3/(3-2p)}\label{eq:Critical_d}
\end{equation}
where $r_{h}$ is the radius of influence (see section 3),
$d_{c}$ is a length scale and $p$ is the power law of the distribution
function ($f(E,t)\sim E^{p}$, see section 3
and \citealt{1976ApJ...209..214B}).

The event rate of EMRIs has been estimated by several studies (e.g.
\citet{Hils1995,1997MNRAS.284..318S,2001CQGra..18.4033F,2002MNRAS.336..373I,2003ApJ...590L..29A,2006ApJ...645L.133H})
but remains rather uncertain, in part because of the slow nature of
the inspiral process, which occurs on many dynamical times (see further
discussion by \citet{2006ApJ...645L.133H}).

Previous studies of EMRI rates in NSCs typically considered only populations
of single-mass SBHs (with mass $m_{\bullet}=10$ M$_{\odot}$), however,
theoretical studies, and the recent GW detection of a merger of two
$\sim30$ M, suggest a potentially wide range of SBH masses. The effect
of such non-trivial SBH population on the rates of GWs from merger
of binary-SBH (detectable by aLIGO) was considered by \citet{Ole+09}.
Here we focus on the implications for a different type of GW sources,
namely inspirals of COs on MBHs, producing EMRIs. Future EMRI detections
could potentially probe the mass function (MF) of inspiralling SBHs.
However, the mutual interactions between SBHs and stars of different
masses could significantly alter their distributions near MBHs, and
thereby the EMRI rate from SBHs of different masses. Therefore, translating
the EMRI MF into the original MF of SBHs requires understanding the
non-trivial evolution and mass segregation processes in NSCs. Finally,
we note that not only the MF of SBHs change their distribution, but
potentially the star-formation history and build-up of the NSC \citep[see][]{2014ApJ...784L..44P,2014ApJ...794..106A,2015ApJ...799..185A,Aha+16}. 

In this letter, we explore for the first time the expected mass-function
of EMRIs (with the main focus on SBHs), and its relation and translation
to the general MF of SBHs. Furthermore, we consider both the rates
from relaxed NSCs, as well the role of the build-up and star-formation
history in affecting the EMRI properties and rates.

\section{Analytic derivation of EMRI rates in relaxed nuclear clusters\label{sec:Analytic-derivation-of}}

In relaxed NSC systems the distribution of different stellar populations
can typically achieve a steady state. Studies of simple stellar populations
(composed of 4 populations: Solar mass main sequence (MS) stars ,
$1.4$ M$_{\odot}$ NSs, $0.6$ M$_{\odot}$ WDs and 10 M$_{\odot}$
SBHs), using Fokker-Planck (FP; see also below), Monte-Carlo or N-body
simulations showed them to distributed with power-law density profiles,
generally consistent with analytic estimates of mass-segregation effects
near MBH by \citet{1977ApJ...216..883B}. Later studies \citep{2009ApJ...697.1861A}
pointed-out that the mass-segregation solution for the steady-state
distribution of stars around a MBH has two branches: a weak-segregation
solution (described by \citealp{1977ApJ...216..883B}) and a different,
strong-segregation solution. They found that their properties depends
on the heavy-to-light stellar mass ratio $M_{H}/M_{L}$ and on the
unbound population number ratio $N_{H}/N_{L}$, through the relaxational
coupling parameter
\begin{equation}
\Delta=4N_{H}M_{H}^{2}/[N_{L}M_{L}^{2}(3+M_{H}/M_{L})],\label{eq:Delta}
\end{equation}
where systems with $\Delta\lesssim1$ reside in the strong mass-segregation
regime. In the strong mass-segregation regime the massive objects
can achieve much stepper density profile compared with the \citet{1977ApJ...216..883B}
weak mass-segregation regime. \citet{2009ApJ...697.1861A} mainly
focused on cases which can be generally divided between small population
of massive objects (${\rm M_{H}\sim}10$ M$_{\odot}$ SBHs) and large
population of low-mass objects (MS stars, NSs and WDs with ${\rm M_{L}}\sim1$${\rm M}{}_{\odot}$)
in the $\Delta<1$ regime. However, a more complex situation arises
when the high-mass population is composed of a range of masses. For
example, in the case of two massive populations (with ${\rm M_{H1},\, M_{H2}}$
and ${\rm N_{H1},\, N_{H2}})$, the calculated $\Delta$ parameter
could be below unity when comparing each of the high-mass populations
to the low-mass one, while $\Delta$ could be above unity when comparing
the two massive-stars populations (i.e. treating ${\rm M_{H2}}$ as
a low mass compared with ${\rm M_{H1}}$). \citet{2009ApJ...698L..64K}
used analytic tools and generalized the derivation for such non-trivial
MF cases. 

Using the above mentioned results, one can find the expected density
profile ($n(r)\propto r^{-\gamma}$), where $\gamma=p+3/2,$ then
for a given population of stars with a given mass in a relaxed NSC,
given by \citep{2009ApJ...698L..64K} and \citet{2009ApJ...697.1861A}:

\begin{equation}
p(m)\backsimeq m/4M_{0}\label{eq:profile}
\end{equation}
where $m$ is the relevant population mass and $M_{0}$ is the weight
average mass. for mass function not strongly dominated by light stars
(negligible flow) the relation is linear: $p\propto m$ \citep{1977ApJ...216..883B,2009ApJ...698L..64K}.
In order to find the number of GW progenitors we then need to integrate
the number of CO GW progenitors (i.e. inside the critical separation,
$a_{GW}$:
\begin{equation}
\Gamma\sim\int^{a_{GW}}r^{-\gamma}d^{3}r\sim a_{GW}^{3/2-p}.\label{eq:a_gw}
\end{equation}
Substituting Eq. \ref{eq:Critical_d} then gives us the inspiral rate
dependence on the the CO mass per unit mass:

\begin{equation}
\Gamma(m)dm\sim a_{GW}^{\frac{3}{2}-\frac{m}{4M_{0}}}=\left(r_{h}\left(\frac{d_{c}}{r_{h}}\right)^{3/(3-\frac{m}{2M_{0}})}\right)^{\frac{3}{2}-\frac{m}{4M_{0}}}dm
\end{equation}
Given the GW detector sensitivity on the inspiraling CO mass, COs
with ${\rm M_{1}>M_{2}}$would be detected to distances larger by
${\rm M_{2}/M_{1}}$ and taking a homogenous universe at sufficiently
large distances one expects a $({\rm M_{2}/M_{1})^{3}}$ enhancement
in the detection rate. Given an intrinsic MF of COs (for simplicity
assuming a power-law distribution; $\xi(m)\propto m^{-\beta}$), we
can now combine all of the above EQs together and relate the CO intrinsic
MF to the MF of detected EMRIs:

\begin{equation}
N(m)\sim\int\left(r_{h}\left(\frac{d_{c}}{r_{h}}\right)^{3/(3-\frac{m}{2M_{0}})}\right)^{\frac{3}{2}-\frac{m}{4M_{0}}}m^{-\beta}dm
\end{equation}
where the last integral can be numerically solved. Given sufficient
number of EMRI detection one can use these relation to derive the
intrinsic MF of SBH.

\section{Numerical calculations using a Fokker-Planck approach \label{sec:EVOLUTION-OF-AN}}

In order to test the analytic results we use a FP approach. Our model
is based on the classic approach of \citet{1976ApJ...209..214B} and
\citet{1977ApJ...216..883B} to the problem, and use our parallelized
FP code (as described in \citealp{2015ApJ...799..185A,Aha+16}). We
simulate the evolution in time, $t,$ of the energy, $E,$ distribution
function (DF) - $f(E,t)$ and the number density of stars in a spherical
system around a MBH. The DF represents the distribution of stars in
central few pcs, and in particular in the range between the Schwarzschild
radius and the radius of influence. To our original code we now added
the treatment of multi-mass cases, following the same equations as
described in \citet{1977ApJ...216..883B} and \citet{2006ApJ...645L.133H}. 

In the following, we briefly recapitulate the main assumptions and
discuss our treatment of GW capture.

\subsection{Fokker-Planck analysis }

The FP model used consists of a time and and energy-dependent, angular
momentum-averaged particle conservation equation. It has the form:

\begin{equation}
\frac{\partial f(E,t)}{\partial t}=-AE^{-\frac{5}{2}}\frac{\partial F}{\partial E}-F_{LC}(E,T)+F_{SF}(E,T)\label{eq:Fokker_Planck}
\end{equation}

where 
\begin{equation}
A=\frac{32\pi^{2}}{3}G^{2}M_{*}^{2}\ln(\Lambda)
\end{equation}

The term $F=F[f(E),E]$ is related to the stellar flow, and plays
an important role in the evolution of the stellar cluster. It presents
the flow of stars in energy space due to two-body relaxation, it is
defined by:

\begin{equation}
F=\int dE'\left(f(E)\frac{\partial f(E')}{\partial E'}-f(E')\frac{\partial f(E)}{\partial E}\right)\left(\max(E,E')\right)^{-\frac{3}{2}}.\label{eq:flow_eq}
\end{equation}

We note that similar to \citet{2006ApJ...645L.133H} work, we neglect
here the effect of resonant relaxation (RR; \citealt{1996NewA....1..149R};\citealt{1998MNRAS.299.1231R}),
which was suggested to increase the EMRI rate by up to an order of
magnitude (\citealt{2006ApJ...645.1152H} and \citet{Merritt2015}.
However, more recent analysis of (\citealt{2016ApJ...820..129B},
who also considered Merritt work, Bar-Or, private communication),
suggest a relatively negligible effect, supporting our approach on
this issue.

\subsection{NSC models}

We followed the evolution of several types of NSCs with a MBH mass
of $4\times10^{6}M_{\odot}$ which can be a representative of a typical
eLISA sources. We first studied primordial NSCs which achieved a steady
state, and did not experience any star formation (SF). Such models
correspond to the same type of models considered previously \citep[e.g.][]{2006ApJ...649...91F,2006ApJ...645L.133H},
besides the difference in stellar population used (see next section;
we also recalculated the same models used by \citealp{2006ApJ...645L.133H},
and verified we get the same results in this case).

For the model with SF we tested two cases: inner SF formation in the
range $0.05-0.5{\rm pc}$ and outer SF between $2-3.5{\rm pc}$. For
each of these two cases we considered two different scenarios of formation
and evolution: 1) in situ formation of MS stars and COs. 2) An initial
pre-existing cusp of MS+CO stars evolved with continuous SF MS+CO
stars (see Table \ref{tab:NSC-moder-and_GW}). In the latter scenario,
we define the initial MS DF within $r_{h}$ in the form of $f(E_{r<r_{h}},t_{0})\propto E^{0.25}$
corresponding to the BW steady-state cusp which has the form of $n\propto r^{-7/4}$.
The density profile at $r_{h}$ is normalized to $4\times10^{4}{\rm pc^{-3}}$,
corresponding to number density at $r_{h}$ in the galactic center
(GC) assuming a mean mass of $1M_{\odot}$ \citep{Genzel2003}. We
also adopt the following parameters taken from the GC values: $\sigma_{\star}=75{\rm km/s}$
and $r_{h}=2{\rm pc}$ \citep{Genzel2003}. We consider the DF at
$r>r_{h}$ to have a Maxwellian distribution: $f(E_{r>r_{h}},t)\propto e^{E/\sigma_{\star}}$
\citep{1976ApJ...209..214B}.

In order to account for SF, following our previous work in \citet{2015ApJ...799..185A},
we added the source term that represents the stars and CO formation
in the form:

\begin{equation}
F_{SF}(E,T)=\frac{\partial}{\partial t}\left(\Pi(E)E_{0}E^{\alpha}\right),\label{eq:SF_sourceT}
\end{equation}
where $\Pi(E)$ is a rectangular function, which boundaries correspond
to the region where new stars are assumed to from; $E_{0}$ is the
source term amplitude; and $F_{SF}$ is a power-law function with
a slope $\alpha$, defining the SF distribution in phase space.

\subsection{Types of Stellar Objects }

Similar to \citealt{2006ApJ...645L.133H} we consider four basic populations.
The first consists of MS stars, assumed here to be of solar mass.
MSs do not contribute to the GW inspiral rate since they are tidally
disrupted before spiraling in, but they do contribute dynamically.
The other populations represent WDs ($M_{WD}=0.6M_{\odot}$), NSs
($M_{NS}=1.4M_{\odot}$), and SBHs. We considered cases which include
stellar populations of $10$ ${\rm M_{\odot}}$SBHs (similar to previous
studies), but also considered cases with both $10$ ${\rm M_{\odot}}$
and $30$ ${\rm M_{\odot}}$SBHs, in order to probe the effect of
more than a single type of massive objects. Considering \citet{Kroupa2001}
work, we used the fraction ratios of the four populations as $C_{MS}:C_{WD}:C_{NS}:C_{BH_{10M_{\odot}}}:C_{BH_{30M_{\odot}}}=0.72:0.26:0.014:2.3\times10^{-3}:2.2\times10^{-4}$,
typical for continuously star-forming populations. The MF of SBHs
is not well understood, however for SBHs up to $\sim40$ ${\rm M_{\odot}}$
theoretical models of direct collapse suggest an almost linear relation
between progenitor mass and final SBH mass \citep{Bel+16}, and we
therefore assume an SBH MF which goes like $\propto m^{-2.1}$ close
to the initial MF of massive stars ($m^{-2.3}$).

\section{RESULTS }

\subsection{Distribution of compact objects populations in NSCs \label{sub:Distribution-of-compact}}

\begin{figure*}
\includegraphics[scale=0.62]{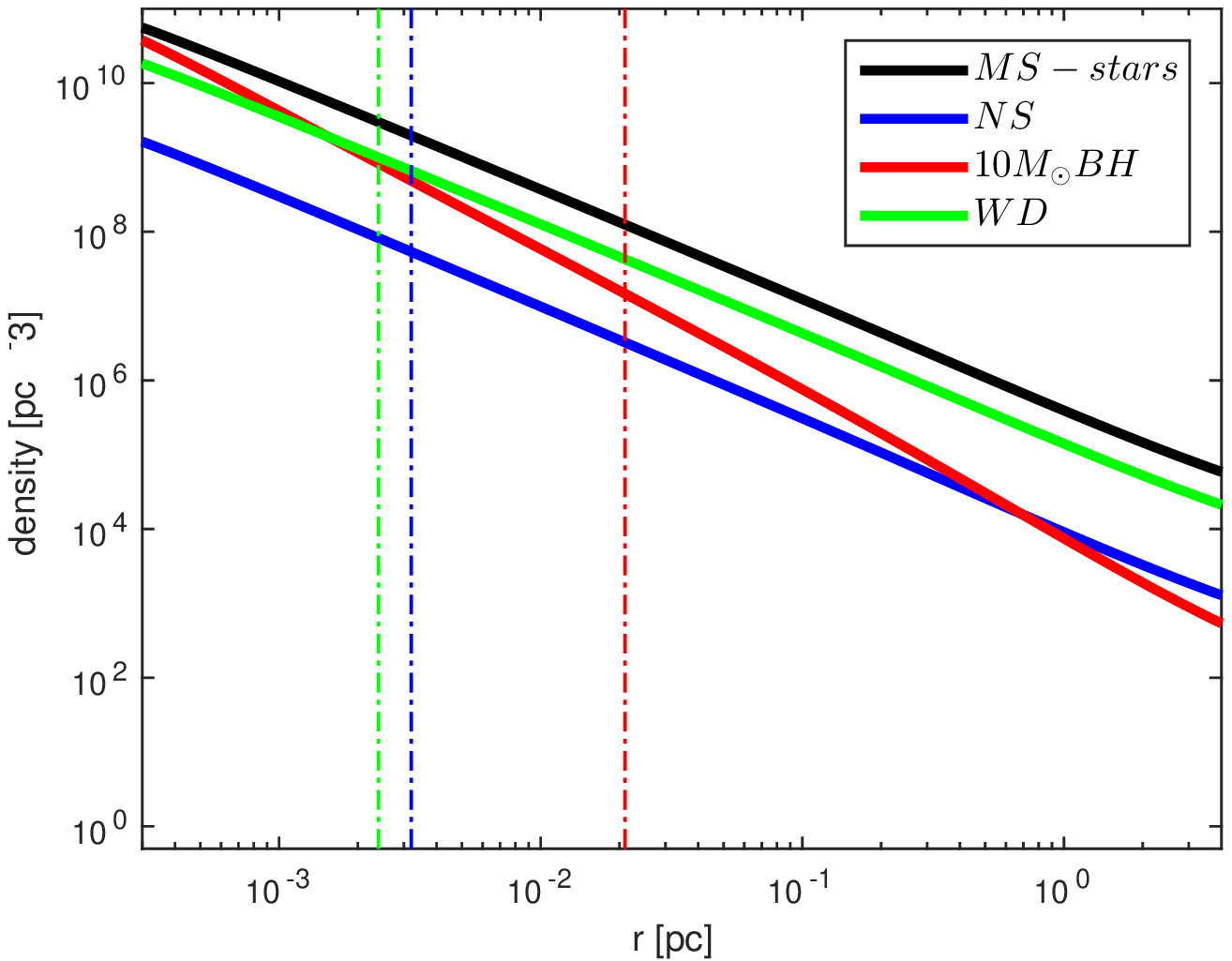}\includegraphics[scale=0.62]{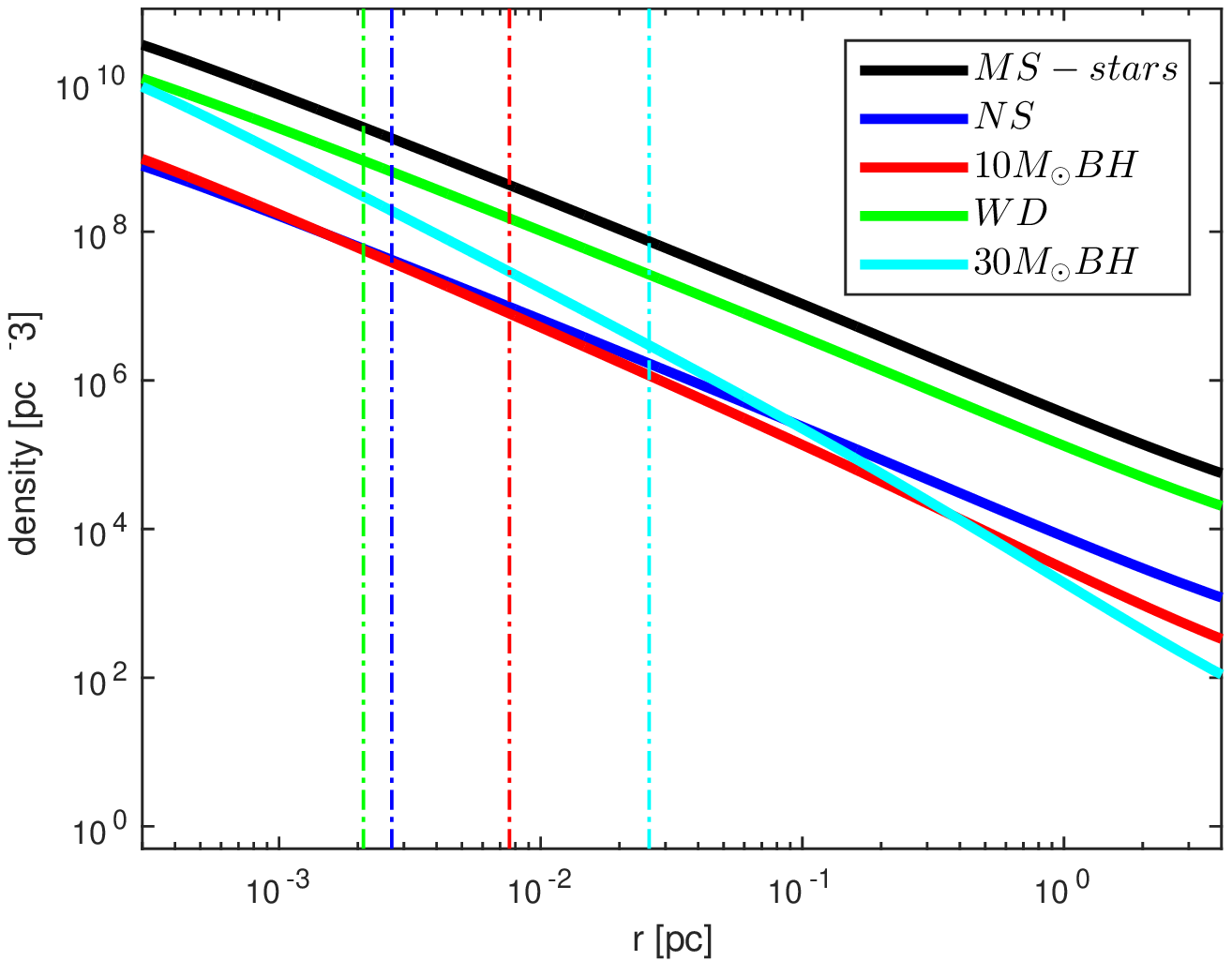}

\protect\caption{Density profile of a 10 Gyr evolved compact stellar objects. Left
panel: pre-existing BW cusp with COs evolved from outer CO formation
at distances $2-3.5{\rm pc}$ from the MBH at the rates of $10^{-4}{\rm yr^{-1}}$
for 4 different stellar populations (MSs, NSs, WDs and $10M_{\odot}$
SBHs). The dashed dot vertical lines mark the maximal semimajor axis
for each CO population where it experiences ``successful inspiral\textquotedblright .
For each population, the colors of the dashed dot lines are corresponding
with the colors of its density profile. The relaxational coupling
parameter for the $10M_{\odot}$ is $\Delta\approx0.87$. The power-laws
for each population within its $a_{GW}$ are $\gamma_{NS}=1.4,\,\gamma_{WD}=1.3$
and $\gamma_{10M_{\odot}SBH}=1.9$. Right panel: similar NSC, with
the addition of stellar population composed of $30M_{\odot}$ SBHs.
The relaxational coupling parameter for the $30M_{\odot}$ is $\Delta\approx0.08$,
and the power-laws are $\gamma_{NS}=1.3,\,\gamma_{WD}=1.3,\,\gamma_{10M_{\odot}SBH}=1.4$
and $\gamma_{30M_{\odot}SBH}=2.1$. \label{fig:3 basic Densities}}
\end{figure*}

\begin{figure*}
\includegraphics[scale=0.62]{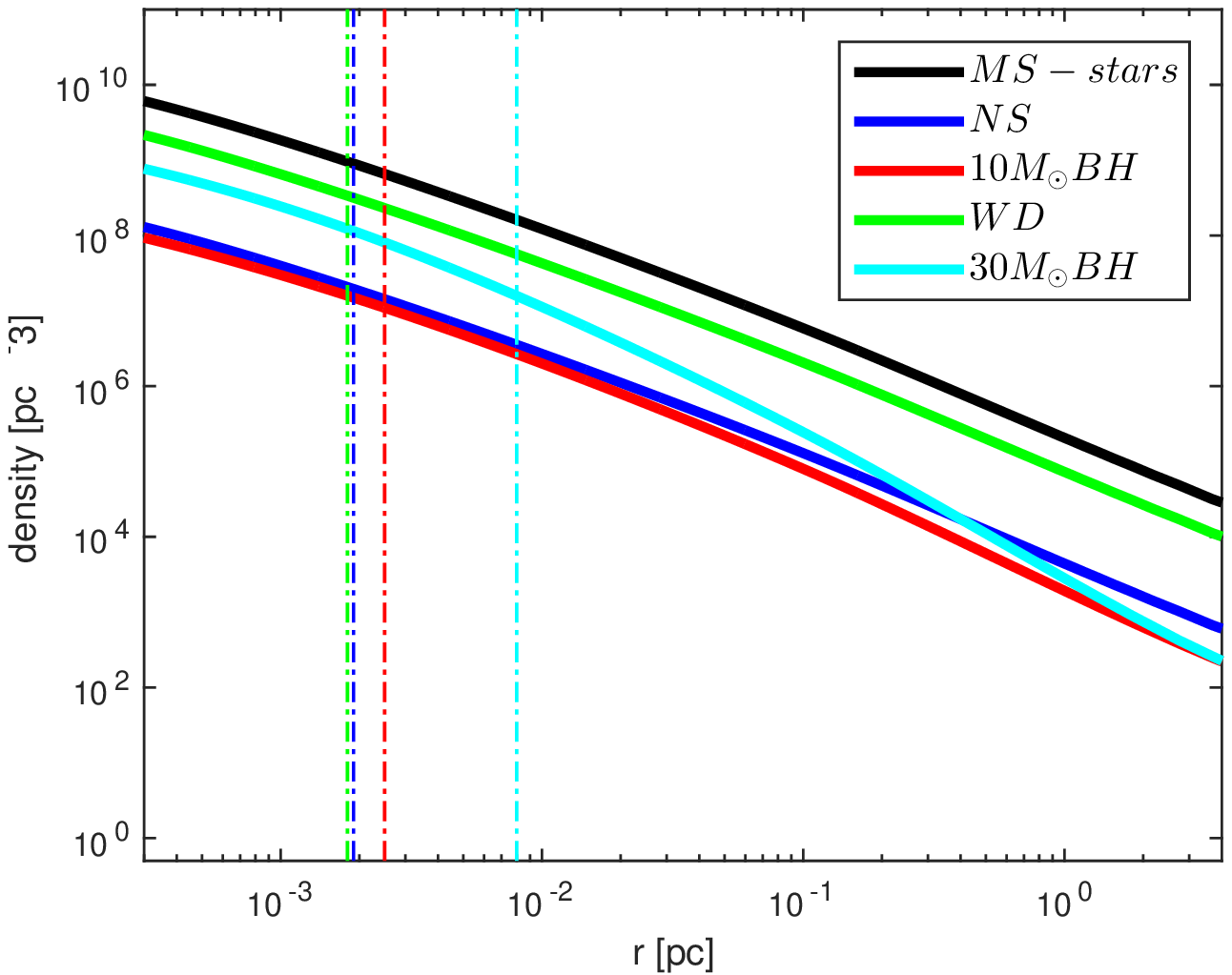}\includegraphics[scale=0.62]{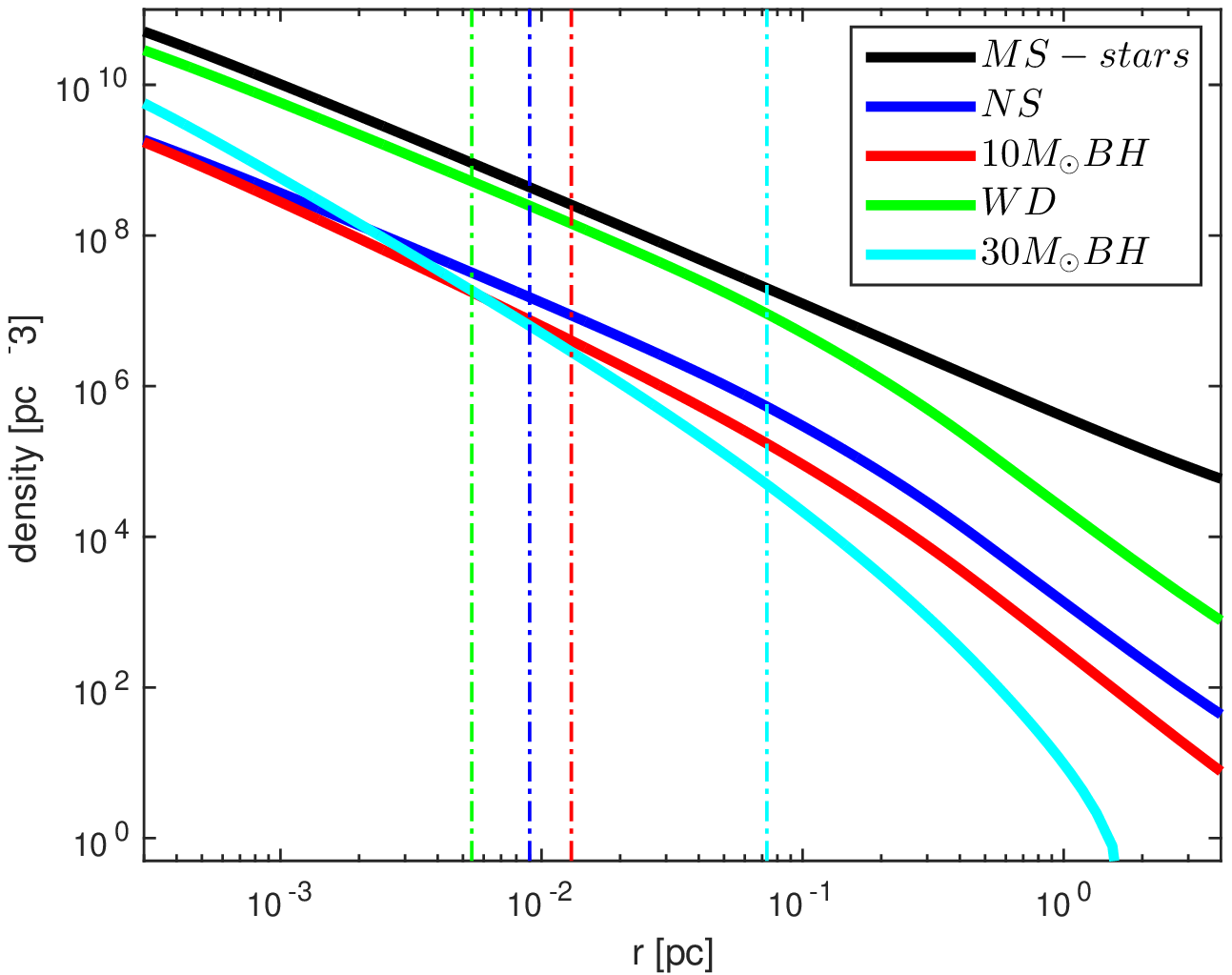}

\protect\caption{Two other GW predicted scenarios. Left panel: Density profile of NSCs
evolved through in situ SF with a rate of $5\times10^{-4}{\rm yr^{-1}}$
(built up NSC) in the outer region ($2-3.5{\rm pc}$). The relaxational
coupling parameter for the $30M_{\odot}$ is $\Delta\approx5.92$.
The power-laws are $\gamma_{NS}=1.0,\,\gamma_{WD}=1.0,\,\gamma_{10M_{\odot}SBH}=1.0$
and $\gamma_{30M_{\odot}SBH}=1.1$. Right panel: CO formation in the
inner region ($0.05-0.5{\rm pc}$) in NSC that evolves from pre-existing
cusp. The relaxational coupling parameter for the $30M_{\odot}$ is
$\Delta\approx0.03$, and the power-laws for each population within
its $a_{GW}$ are $\gamma_{NS}=1.3,\,\gamma_{WD}=1.3$, $\gamma_{10M_{\odot}SBH}=1.5$
and $\gamma_{30M_{\odot}SBH}=2.3$ \label{outer_inner_SF_higher_lower}}
\end{figure*}

We followed the evolution of the studied NSCs for 10Gyr corresponding
to Hubble timescale. Note that this timescale is also comparable with
the relaxation time for a GC-like NSC, but likely lower that the relaxation
time for NSCs hosting more massive MBH. We present the density profile
of the studied scenarios in Figs \ref{fig:3 basic Densities}, and
\ref{outer_inner_SF_higher_lower}, where we focus the central $0.01{\rm pc}$
region to predict the GW rates ($10^{-3}$ for WDs and NSs). We integrate
the density profile in the mentioned region and obtain the number
and derive the rates of inspiralling CO. We summarize our results
in Table \ref{tab:NSC-moder-and_GW}. For comparison we obtained results
from simulations similar to those of \citeauthor{2006ApJ...645L.133H}
(2006; Fig. \ref{fig:3 basic Densities}; i.e. a steady state NSC
with no SF). 

As expected from the effects of strong mass-segregation the density
profile of the SBHs in most scenario is much steeper than the $\sim$Solar
mass MS/CO stars. Consequently, the rates of EMRI of SBHs is the highest
compared with NSs and WDs. In the inner regions of the NSC, SBHs can
become the most frequent stellar species, where the exact region where
the dominate depends on the specific model explored. 

The distribution of the $10M_{\odot}$ SBHs in NSCs that include $30M_{\odot}$
SBHs is very similar to the NS distribution, where in NSCs without
$30M_{\odot}$ SBHs, the distribution of $10M_{\odot}$ has the steeper
slope compared to the other populations. In other words, in the absence
of $30M_{\odot}$ SBHs, the $10M_{\odot}$ SBHs distribution is border-line
between the weak and strong mass-segregation regimes. The existence
of the $30M_{\odot}$ population quenches the effects of strong mass-segregation
on the $10M_{\odot}$, and only the $30M_{\odot}$ SBHs are strongly
segregated. 

The obtained slope of the $10M_{\odot}$ SBH number density in the
outer-SF scenario is $\gamma_{10M_{\odot}SBH}^{out}\approx1.9$, where
in the scenario that includes the $30M_{\odot}$ SBHs it decreases
to $1.42$, and the slope of the more massive SBHs is $\gamma_{30M_{\odot}SBH}^{out}\approx1.9$
(Fig. \ref{fig:3 basic Densities}). The highest slope ($\gamma_{30M_{\odot}SBH}^{out}\approx2.3$)
is obtained for the $30M_{\odot}$ SBH population in NSCs, evolved
from pre-existing stellar cusp, that experience inner-SF. The slopes
calculated in our numerical simulations are consistent with the analytical
study of \citet{2009ApJ...698L..64K}. \textcolor{black}{We emphasize
that we also tested EMRI rates model with a range of SBH masses in
order to verify our }Analytic derivation of the rates presented in
section 2, and found comparable results.

\subsection{EMRI rates}

Following the obtained NSC stellar distribution, we integrated the
number of COs up to the critical separation in order to quantify the
number of expected EMRIs. We summarize the results in Table \ref{tab:NSC-moder-and_GW}.
The highest EMRIs are obtained in NSCs that evolve from a pre-existing
cusp that also experiences strong SF in the outer region. The lowest
overall rates are obtained with inner-SF that builds up the NSC. For
comparison with previous studies, the last row in the table is based
on \citet{2006ApJ...645L.133H} as described in \ref{sub:Distribution-of-compact}. 

\begin{table*}[t]
\begin{tabular}{|>{\centering}p{3cm}||>{\centering}p{2cm}|c|c|c|c|}
\hline 
{\footnotesize{}scenario} & {\footnotesize{}CO formation rate (${\rm yr^{-1}}$)} & {\footnotesize{}$\Gamma_{NS}$ (${\rm Gyr^{-1}}$) } & {\footnotesize{}$\Gamma_{WD}$ (${\rm Gyr^{-1}}$) } & {\footnotesize{}$\Gamma_{SBH_{10M\odot}}$ (${\rm Gyr^{-1}}$) } & {\footnotesize{}$\Gamma_{SBH_{30M\odot}}$ (${\rm Gyr^{-1}}$) }\tabularnewline
\hline 
\hline 
{\footnotesize{}pre-existing cusp outer formation without $30M_{\odot}$
SBH} & \multirow{5}{2cm}{{\footnotesize{}$10$$^{-4}$}} & {\footnotesize{}14} & {\footnotesize{}79} & {\footnotesize{}232} & {\footnotesize{}-}\tabularnewline
\cline{1-1} \cline{3-6} 
{\footnotesize{}pre-existing cusp outer formation } &  & {\footnotesize{}11} & {\footnotesize{}71} & {\footnotesize{}92} & {\footnotesize{}265}\tabularnewline
\cline{1-1} \cline{3-6} 
{\footnotesize{}pre-existing cusp inner formation } &  & {\footnotesize{}15} & {\footnotesize{}73} & {\footnotesize{}87} & {\footnotesize{}252}\tabularnewline
\cline{1-1} \cline{3-6} 
{\footnotesize{}outer in-situ formation} &  & {\footnotesize{}7} & {\footnotesize{}22} & {\footnotesize{}58} & {\footnotesize{}273}\tabularnewline
\cline{1-1} \cline{3-6} 
{\footnotesize{}inner in-situ formation} &  & {\footnotesize{}39} & {\footnotesize{}62} & {\footnotesize{}74} & {\footnotesize{}312}\tabularnewline
\hline 
{\footnotesize{}outer in-situ formation} & \multirow{3}{2cm}{{\footnotesize{}$5\times10$$^{-4}$}} & {\footnotesize{}32} & {\footnotesize{}112} & {\footnotesize{}62} & {\footnotesize{}288}\tabularnewline
\cline{1-1} \cline{3-6} 
{\footnotesize{}inner in-situ formation} &  & {\footnotesize{}45} & {\footnotesize{}68} & {\footnotesize{}85} & {\footnotesize{}301}\tabularnewline
\cline{1-1} \cline{3-6} 
{\footnotesize{}pre-existing cusp inner formation} &  & {\footnotesize{}8} & {\footnotesize{}67} & {\footnotesize{}15} & {\footnotesize{}97}\tabularnewline
\hline 
{\footnotesize{}relaxed cusp} & {\footnotesize{}-} & {\footnotesize{}7} & {\footnotesize{}73} & {\footnotesize{}89} & {\footnotesize{}273}\tabularnewline
\hline 
{\footnotesize{}relaxed cusp without $30M_{\odot}$ SBH (based on
\citealt{2006ApJ...645L.133H})} & {\footnotesize{}-} & {\footnotesize{}6} & {\footnotesize{}33} & {\footnotesize{}252} & {\footnotesize{}-}\tabularnewline
\hline 
\end{tabular}

\protect\caption{NSC model and the predicted GW rates\label{tab:NSC-moder-and_GW}}
\end{table*}

\section{SUMMARY}

In this work we studied the rates of GWs from extreme-mass ratio inspirals,
but considered two novel aspects which were little, or not considered
before in this context. We study (1) the impact of a wide mass-spectrum
SBHs (as suggested by theoretical work and the recent detections of
high mass SBHs by aLIGO; \citealp{2016PhRvL.116f1102A,Bel+16}) on
EMRI rates, and (2) the role of in-situ formation of stars and COs
during the build-up and/or formation of nuclear stellar clusters. 

Our main findings are as follows:
\begin{enumerate}
\item We find that strong mass-segregation produces a steep power-law density
profile for the most massive SBHs, but at the same time quenches the
migration of less massive SBHs close to the MBH. We quantify this
effect and provide a translation between the intrinsic mass function
of CO, in particular SBHs, and the (future) observable mass function
of the EMRI GW sources, showing the latter to be strongly biased towards
high mass COs. 
\item We find that SF plays a relatively small role in eventually determining
the EMRI rates for SBHs (and its mass function), since most of them
migrate close to the MBH through mass-segregation rather than form
in-situ. However, the rate of EMRI of neutron stars can be significantly
increased when they form in-situ close to the MBH. In this latter
case neutron stars can inspiral before relaxation processes significantly
segregates them outwards farther away from the MBH. A reverse but
weaker effect of decreasing the EMRI rates from neutron stars and
white dwarfs occurs when star-formation proceeds far from the MBH.
In this case the mass segregation processes due to the SBHs somewhat
quench the newly formed NSs/WDs from diffusing into the inner regions,
and lower their EMRI rates. 
\end{enumerate}

\acknowledgements{}

We would like to thank Clovis Hopman the use of the basic components
in his FP code for developing the FP code used in our simulations.
We acknowledge support from the I-CORE Program of the Planning and
Budgeting Committee and The Israel Science Foundation grant 1829/12,
as well support form the Asher Space Research Institute in the Technion. 

\bibliographystyle{plainnat}

\begin{thebibliography}{30}
\providecommand{\natexlab}[1]{#1}
\providecommand{\url}[1]{\texttt{#1}}
\expandafter\ifx\csname urlstyle\endcsname\relax
  \providecommand{\doi}[1]{doi: #1}\else
  \providecommand{\doi}{doi: \begingroup \urlstyle{rm}\Url}\fi

\bibitem[{Abbott} et~al.(2016){Abbott}, {Abbott}, {Abbott}, {Abernathy},
  {Acernese}, {Ackley}, {Adams}, {Adams}, {Addesso}, {Adhikari}, and
  et~al.]{2016PhRvL.116f1102A}
B.~P. {Abbott}, R.~{Abbott}, T.~D. {Abbott}, M.~R. {Abernathy}, F.~{Acernese},
  K.~{Ackley}, C.~{Adams}, T.~{Adams}, P.~{Addesso}, R.~X. {Adhikari}, and
  et~al.
\newblock {Observation of Gravitational Waves from a Binary Black Hole Merger}.
\newblock \emph{Physical Review Letters}, 116\penalty0 (6):\penalty0 061102,
  February 2016.
\newblock \doi{10.1103/PhysRevLett.116.061102}.

\bibitem[{Aharon} and {Perets}(2015)]{2015ApJ...799..185A}
D.~{Aharon} and H.~B. {Perets}.
\newblock {Formation and Evolution of Nuclear Star Clusters with In Situ Star
  Formation: Nuclear Cores and Age Segregation}.
\newblock \emph{\apj}, 799:\penalty0 185, February 2015.
\newblock \doi{10.1088/0004-637X/799/2/185}.

\bibitem[{Aharon} et~al.(2016){Aharon}, {Mastrobuono Battisti}, and
  {Perets}]{Aha+16}
D.~{Aharon}, A.~{Mastrobuono Battisti}, and H.~B. {Perets}.
\newblock {The History of Tidal Disruption Events in Galactic Nuclei}.
\newblock \emph{\apj}, 823:\penalty0 137, June 2016.
\newblock \doi{10.3847/0004-637X/823/2/137}.

\bibitem[{Alexander} and {Hopman}(2003)]{2003ApJ...590L..29A}
T.~{Alexander} and C.~{Hopman}.
\newblock {Orbital In-spiral into a Massive Black Hole in a Galactic Center}.
\newblock \emph{\apjl}, 590:\penalty0 L29--L32, June 2003.
\newblock \doi{10.1086/376672}.

\bibitem[{Alexander} and {Hopman}(2009)]{2009ApJ...697.1861A}
T.~{Alexander} and C.~{Hopman}.
\newblock {Strong Mass Segregation Around a Massive Black Hole}.
\newblock \emph{\apj}, 697:\penalty0 1861--1869, June 2009.
\newblock \doi{10.1088/0004-637X/697/2/1861}.

\bibitem[{Amaro-Seoane} and {Preto}(2011)]{Ama+11}
P.~{Amaro-Seoane} and M.~{Preto}.
\newblock {The impact of realistic models of mass segregation on the event rate
  of extreme-mass ratio inspirals and cusp re-growth}.
\newblock \emph{Classical and Quantum Gravity}, 28\penalty0 (9):\penalty0
  094017, May 2011.
\newblock \doi{10.1088/0264-9381/28/9/094017}.

\bibitem[{Antonini}(2014)]{2014ApJ...794..106A}
F.~{Antonini}.
\newblock {On the Distribution of Stellar Remnants around Massive Black Holes:
  Slow Mass Segregation, Star Cluster Inspirals, and Correlated Orbits}.
\newblock \emph{\apj}, 794:\penalty0 106, October 2014.
\newblock \doi{10.1088/0004-637X/794/2/106}.

\bibitem[{Bahcall} and {Wolf}(1976)]{1976ApJ...209..214B}
J.~N. {Bahcall} and R.~A. {Wolf}.
\newblock {Star distribution around a massive black hole in a globular
  cluster}.
\newblock \emph{\apj}, 209:\penalty0 214--232, October 1976.
\newblock \doi{10.1086/154711}.

\bibitem[{Bahcall} and {Wolf}(1977)]{1977ApJ...216..883B}
J.~N. {Bahcall} and R.~A. {Wolf}.
\newblock {The star distribution around a massive black hole in a globular
  cluster. II Unequal star masses}.
\newblock \emph{\apj}, 216:\penalty0 883--907, September 1977.
\newblock \doi{10.1086/155534}.

\bibitem[{Bar-Or} and {Alexander}(2016)]{2016ApJ...820..129B}
B.~{Bar-Or} and T.~{Alexander}.
\newblock {Steady-state Relativistic Stellar Dynamics Around a Massive Black
  hole}.
\newblock \emph{\apj}, 820:\penalty0 129, April 2016.
\newblock \doi{10.3847/0004-637X/820/2/129}.

\bibitem[{Belczynski} et~al.(2016){Belczynski}, {Heger}, {Gladysz}, {Ruiter},
  {Woosley}, {Wiktorowicz}, {Chen}, {Bulik}, {O'Shaughnesy}, {Holz}, {Fryer},
  and {Berti}]{Bel+16}
K.~{Belczynski}, A.~{Heger}, W.~{Gladysz}, A.~J. {Ruiter}, S.~{Woosley},
  G.~{Wiktorowicz}, H.-Y. {Chen}, T.~{Bulik}, R.~{O'Shaughnesy}, D.~E. {Holz},
  C.~L. {Fryer}, and E.~{Berti}.
\newblock {The Effect of Pair-Instability Mass Loss on Black Hole Mergers}.
\newblock \emph{ArXiv e-prints}, July 2016.

\bibitem[{Eisenhauer} et~al.(2005){Eisenhauer}, {Genzel}, {Alexander},
  {Abuter}, {Paumard}, {Ott}, {Gilbert}, {Gillessen}, {Horrobin}, {Trippe},
  {Bonnet}, {Dumas}, {Hubin}, {Kaufer}, {Kissler-Patig}, {Monnet},
  {Str{\"o}bele}, {Szeifert}, {Eckart}, {Sch{\"o}del}, and
  {Zucker}]{Eisenhauer2005}
F.~{Eisenhauer}, R.~{Genzel}, T.~{Alexander}, R.~{Abuter}, T.~{Paumard},
  T.~{Ott}, A.~{Gilbert}, S.~{Gillessen}, M.~{Horrobin}, S.~{Trippe},
  H.~{Bonnet}, C.~{Dumas}, N.~{Hubin}, A.~{Kaufer}, M.~{Kissler-Patig},
  G.~{Monnet}, S.~{Str{\"o}bele}, T.~{Szeifert}, A.~{Eckart}, R.~{Sch{\"o}del},
  and S.~{Zucker}.
\newblock {SINFONI in the Galactic Center: Young Stars and Infrared Flares in
  the Central Light-Month}.
\newblock \emph{\apj}, 628:\penalty0 246--259, July 2005.
\newblock \doi{10.1086/430667}.

\bibitem[{Freitag}(2001)]{2001CQGra..18.4033F}
M.~{Freitag}.
\newblock {Monte Carlo cluster simulations to determine the rate of compact
  star inspiralling to a central galactic black hole}.
\newblock \emph{Classical and Quantum Gravity}, 18:\penalty0 4033--4038,
  October 2001.
\newblock \doi{10.1088/0264-9381/18/19/309}.

\bibitem[{Freitag} et~al.(2006){Freitag}, {Amaro-Seoane}, and
  {Kalogera}]{2006ApJ...649...91F}
M.~{Freitag}, P.~{Amaro-Seoane}, and V.~{Kalogera}.
\newblock {Stellar Remnants in Galactic Nuclei: Mass Segregation}.
\newblock \emph{\apj}, 649:\penalty0 91--117, September 2006.
\newblock \doi{10.1086/506193}.

\bibitem[{Genzel} et~al.(2003){Genzel}, {Sch{\"o}del}, {Ott}, {Eisenhauer},
  {Hofmann}, {Lehnert}, {Eckart}, {Alexander}, {Sternberg}, {Lenzen},
  {Cl{\'e}net}, {Lacombe}, {Rouan}, {Renzini}, and
  {Tacconi-Garman}]{Genzel2003}
R.~{Genzel}, R.~{Sch{\"o}del}, T.~{Ott}, F.~{Eisenhauer}, R.~{Hofmann},
  M.~{Lehnert}, A.~{Eckart}, T.~{Alexander}, A.~{Sternberg}, R.~{Lenzen},
  Y.~{Cl{\'e}net}, F.~{Lacombe}, D.~{Rouan}, A.~{Renzini}, and L.~E.
  {Tacconi-Garman}.
\newblock {The Stellar Cusp around the Supermassive Black Hole in the Galactic
  Center}.
\newblock \emph{\apj}, 594:\penalty0 812--832, September 2003.
\newblock \doi{10.1086/377127}.

\bibitem[{Ghez} et~al.(2005){Ghez}, {Salim}, {Hornstein}, {Tanner}, {Lu},
  {Morris}, {Becklin}, and {Duch{\^e}ne}]{2005ApJ...620..744G}
A.~M. {Ghez}, S.~{Salim}, S.~D. {Hornstein}, A.~{Tanner}, J.~R. {Lu},
  M.~{Morris}, E.~E. {Becklin}, and G.~{Duch{\^e}ne}.
\newblock {Stellar Orbits around the Galactic Center Black Hole}.
\newblock \emph{\apj}, 620:\penalty0 744--757, February 2005.
\newblock \doi{10.1086/427175}.

\bibitem[{Hils} and {Bender}(1995)]{Hils1995}
D.~{Hils} and P.~L. {Bender}.
\newblock {Gradual approach to coalescence for compact stars orbiting massive
  black holes}.
\newblock \emph{\apjl}, 445:\penalty0 L7--L10, May 1995.
\newblock \doi{10.1086/187876}.

\bibitem[{Hopman} and {Alexander}(2005)]{2005ApJ...629..362H}
C.~{Hopman} and T.~{Alexander}.
\newblock {The Orbital Statistics of Stellar Inspiral and Relaxation near a
  Massive Black Hole: Characterizing Gravitational Wave Sources}.
\newblock \emph{\apj}, 629:\penalty0 362--372, August 2005.
\newblock \doi{10.1086/431475}.

\bibitem[{Hopman} and {Alexander}(2006{\natexlab{a}})]{2006ApJ...645.1152H}
C.~{Hopman} and T.~{Alexander}.
\newblock {Resonant Relaxation near a Massive Black Hole: The Stellar
  Distribution and Gravitational Wave Sources}.
\newblock \emph{\apj}, 645:\penalty0 1152--1163, July 2006{\natexlab{a}}.
\newblock \doi{10.1086/504400}.

\bibitem[{Hopman} and {Alexander}(2006{\natexlab{b}})]{2006ApJ...645L.133H}
C.~{Hopman} and T.~{Alexander}.
\newblock {The Effect of Mass Segregation on Gravitational Wave Sources near
  Massive Black Holes}.
\newblock \emph{\apjl}, 645:\penalty0 L133--L136, July 2006{\natexlab{b}}.
\newblock \doi{10.1086/506273}.

\bibitem[{Ivanov}(2002)]{2002MNRAS.336..373I}
P.~B. {Ivanov}.
\newblock {On the formation rate of close binaries consisting of a
  super-massive black hole and a white dwarf}.
\newblock \emph{\mnras}, 336:\penalty0 373--381, October 2002.
\newblock \doi{10.1046/j.1365-8711.2002.05733.x}.

\bibitem[{Keshet} et~al.(2009){Keshet}, {Hopman}, and
  {Alexander}]{2009ApJ...698L..64K}
U.~{Keshet}, C.~{Hopman}, and T.~{Alexander}.
\newblock {Analytic Study of Mass Segregation Around a Massive Black Hole}.
\newblock \emph{\apjl}, 698:\penalty0 L64--L67, June 2009.
\newblock \doi{10.1088/0004-637X/698/1/L64}.

\bibitem[{Kroupa}(2001)]{Kroupa2001}
P.~{Kroupa}.
\newblock {On the variation of the initial mass function}.
\newblock \emph{\mnras}, 322:\penalty0 231--246, April 2001.
\newblock \doi{10.1046/j.1365-8711.2001.04022.x}.

\bibitem[{Merritt}(2015)]{Merritt2015}
D.~{Merritt}.
\newblock {Gravitational Encounters and the Evolution of Galactic Nuclei. IV.
  Captures Mediated by Gravitational-wave Energy Loss}.
\newblock \emph{\apj}, 814:\penalty0 57, November 2015.
\newblock \doi{10.1088/0004-637X/814/1/57}.

\bibitem[{O'Leary} et~al.(2009){O'Leary}, {Kocsis}, and {Loeb}]{Ole+09}
R.~M. {O'Leary}, B.~{Kocsis}, and A.~{Loeb}.
\newblock {Gravitational waves from scattering of stellar-mass black holes in
  galactic nuclei}.
\newblock \emph{\mnras}, 395:\penalty0 2127--2146, June 2009.
\newblock \doi{10.1111/j.1365-2966.2009.14653.x}.

\bibitem[{Perets} and {Mastrobuono-Battisti}(2014)]{2014ApJ...784L..44P}
H.~B. {Perets} and A.~{Mastrobuono-Battisti}.
\newblock {Age and Mass Segregation of Multiple Stellar Populations in Galactic
  Nuclei and their Observational Signatures}.
\newblock \emph{\apjl}, 784:\penalty0 L44, April 2014.
\newblock \doi{10.1088/2041-8205/784/2/L44}.

\bibitem[{Preto} and {Amaro-Seoane}(2010)]{Preto2010}
M.~{Preto} and P.~{Amaro-Seoane}.
\newblock {On Strong Mass Segregation Around a Massive Black Hole: Implications
  for Lower-Frequency Gravitational-Wave Astrophysics}.
\newblock \emph{\apjl}, 708:\penalty0 L42--L46, January 2010.
\newblock \doi{10.1088/2041-8205/708/1/L42}.

\bibitem[{Rauch} and {Ingalls}(1998)]{1998MNRAS.299.1231R}
K.~P. {Rauch} and B.~{Ingalls}.
\newblock {Resonant tidal disruption in galactic nuclei}.
\newblock \emph{\mnras}, 299:\penalty0 1231--1241, October 1998.
\newblock \doi{10.1046/j.1365-8711.1998.01889.x}.

\bibitem[{Rauch} and {Tremaine}(1996)]{1996NewA....1..149R}
K.~P. {Rauch} and S.~{Tremaine}.
\newblock {Resonant relaxation in stellar systems}.
\newblock \emph{New Astronomy}, 1:\penalty0 149--170, October 1996.
\newblock \doi{10.1016/S1384-1076(96)00012-7}.

\bibitem[{Sigurdsson} and {Rees}(1997)]{1997MNRAS.284..318S}
S.~{Sigurdsson} and M.~J. {Rees}.
\newblock {Capture of stellar mass compact objects by massive black holes in
  galactic cusps}.
\newblock \emph{\mnras}, 284:\penalty0 318--326, January 1997.
\newblock \doi{10.1093/mnras/284.2.318}.

\end{thebibliography}

\end{document}